\documentclass[a4paper,table,11pt]{article}
\usepackage{pos}
\usepackage{amsmath}
\usepackage{xcolor}
\usepackage{url}
\usepackage{graphicx}
\usepackage{booktabs, multirow}
\usepackage{graphbox}
\usepackage{esvect}
\usepackage{axodraw2}
\usepackage{enumitem}
\setlist{nosep}

\newcommand{\brk}[1]{(#1)}
\newcommand{\lrbrk}[1]{\left(#1\right)}
\newcommand{\bigbrk}[1]{\bigl(#1\bigr)}

\newcommand{\lrsbrk}[1]{\left[#1\right]}

\newcommand{\brc}[1]{\{#1\}}

\newcommand{\bigbrc}[1]{\bigl\{#1\bigr\}}

\newcommand{\vev}[1]{\langle #1\rangle}

\newcommand{\bra}[1]{\langle #1|}
\newcommand{\ket}[1]{|#1\rangle}

\newcommand{\dd}{\mathrm{d}}
\newcommand{\dlog}{\dd \log}

\newcommand{\defas}{:=}

\newcommand{\twist}{u}

\newcommand{\Integers}{\mathbb{Z}}

\newcommand{\Complex}{\mathbb{C}}

\newcommand{\res}[1]{\mathop{\mathrm{Res}}_{#1}}

\newcommand{\Dim}{\mathop{\mathrm{dim}}}
\newcommand{\cp}{\mathrm{\Complex P}}

\newcommand{\cJ}{\ensuremath{\mathcal{J}}}

\newcommand{\FI}{\mathcal{I}}
\newcommand{\Loops}{\ell}
\newcommand{\dims}{d}

\newcommand{\contour}{\mathcal{C}}
\newcommand{\Baikov}{\mathcal{B}}
\newcommand{\ccl}{\varphi}
\newcommand{\ccr}{\psi}
\newcommand{\cohom}{\mathbb{H}}
\newcommand{\rank}{r}
\newcommand{\lbasis}{e}
\newcommand{\rbasis}{h}
\newcommand{\Cmat}{C}

\newcommand{\Poles}{\mathbb{P}}
\newcommand{\potential}{f}
\newcommand{\Max}{\mathrm{max}}
\newcommand{\Min}{\mathrm{min}}
\newcommand{\Pfaff}{P}
\newcommand{\aux}{^{\mathrm{aux}}}
\newcommand{\mzero}{\gr{\cdot}}
\newcommand{\transpose}{^\mathrm{T}}
\newcommand{\eps}{\varepsilon}
\newcommand{\imi}{\mathrm{i}}
\newcommand{\Id}{\mathrm{id}}
\newcommand{\IdOp}{\mathbb{I}}
\newcommand{\subc}{_{\mathrm{c}}}

\newcommand{\Heaviside}{\theta}

\makeatletter
    \newcommand*\bigcdot{\mathpalette\bigcdot@{1}}
    \newcommand*\smtimes{\mathpalette\smtimes@{.7}}
    \newcommand*\bigcdot@[2]{\mathbin{\vcenter{\hbox{\scalebox{#2}{$\m@th#1\bullet$}}}}}
    \newcommand*\smtimes@[2]{\mathbin{\vcenter{\hbox{\scalebox{#2}{$\m@th#1\times$}}}}}
\makeatother

\newcommand{\soft}[1]{\textsc{#1}}

\newcommand{\namedref}[2]{\hyperref[#2]{#1~\ref*{#2}}}
\newcommand{\secref}[1]{\namedref{Section}{#1}}

\makeatletter
\def\mr@ignsp#1 {\ifx\:#1\@empty\else #1\expandafter\mr@ignsp\fi}%
\newcommand{\multiref}[1]{\begingroup%\let\protect\string%
\xdef\mr@no@sparg{\expandafter\mr@ignsp#1 \: }%
\def\mr@comma{}%
\@for\mr@refs:=\mr@no@sparg\do{\mr@comma\def\mr@comma{,\,}\ref{\mr@refs}}%
\endgroup}
\makeatother
\renewcommand{\eqref}[1]{(\multiref{#1})}

\newcommand{\Res}{{\rm Res}}

\makeatletter
\newlength{\apb@width}
\newcommand{\autoparbox}[2][c]{\settowidth{\apb@width}{#2}\parbox[#1]{\apb@width}{#2}}

\makeatother

\definecolor{green1}{HTML}{244819}
\definecolor{cyan1}{HTML}{37cdaa}
\definecolor{blue1}{HTML}{5d7ac4}
\definecolor{red1}{HTML}{d0482a}
\definecolor{purple1}{HTML}{845ea8}
\definecolor{orange1}{HTML}{e07229}
\definecolor{gr}{gray}{0.5}
\definecolor{gr1}{gray}{0.9}
\newcommand{\gr}[1]{{\color{gr}#1}}

\title{
    Recent progress in intersection theory for Feynman integrals decomposition
}
\author*[a,b]{Vsevolod Chestnov}
\newcommand{\unipd}{Dipartimento di Fisica e Astronomia, Universit\`a degli Studi di Padova,
Via Marzolo 8, I-35131 Padova, Italy.}
\newcommand{\pdinfn}{INFN, Sezione di Padova,
Via Marzolo 8, I-35131 Padova, Italy.}

\affiliation[a]{\unipd}
\affiliation[b]{\pdinfn}
\emailAdd{\\vsevolod.chestnov@pd.infn.it}
\abstract{
    High precision calculations in perturbative QFT often require evaluation of big
    collection of Feynman integrals. Complexity of this task can be greatly reduced
    via the usage of linear identities among Feynman integrals.  Based on
    mathematical theory of intersection numbers, recently a new method for
    derivation of such identities and decomposition of Feynman integrals was
    introduced and applied to many non-trivial examples.

    In this note based on~\cite{Chestnov:2022alh} we discuss the latest
    developments in algorithms for the evaluation of intersection numbers, and
    their application to the reduction of Feynman integrals.
}

\FullConference{%
  Loops and Legs in Quantum Field Theory - LL2022,\\
  25-30 April, 2022\\
  Ettal, Germany
}

\def\boxx{
\raisebox{-8pt}{
\hspace{-5pt}
\begin{axopicture}{(20,20)(0,0)}
    \SetWidth{0.4}
    \Line(5,5)(15,5)
    \Line(15,5)(15,15)
    \Line(15,15)(5,15)
    \Line(5,15)(5,5)
    \Line(0,0)(5,5)
    \Line(20,0)(15,5)
    \Line(20,20)(15,15)
    \Line(0,20)(5,15)
\end{axopicture}
}}
\def\boxd{
\raisebox{-8pt}{
\hspace{-5pt}
\begin{axopicture}{(20,20)(0,0)}
    \SetWidth{0.8}
    \Line(5,5)(15,5)
    \Line(15,5)(15,15)
    \Line(15,15)(5,15)
    \Line(5,15)(5,5)
    \Line(0,0)(5,5)
    \Line(20,0)(15,5)
    \Line(20,20)(15,15)
    \Line(0,20)(5,15)
    \Vertex(10,5){2}
\end{axopicture}
}}
\def\tbub{
\raisebox{-8pt}{
\hspace{-5pt}
\begin{axopicture}{(20,20)(0,0)}
    \SetWidth{0.8}
    \ECirc(10,10){6}
    \Vertex(16,10){2}
    \Line(5,20)(10,16)
    \Line(15,20)(10,16)
    \Line(5,0)(10,4)
    \Line(15,0)(10,4)
\end{axopicture}
\hspace{-2pt}
}}
\def\sbub{
\raisebox{-8pt}{
\hspace{-5pt}
\begin{axopicture}{(20,20)(0,0)}
    \SetWidth{0.8}
    \ECirc(10,10){6}
    \Vertex(10,16){2}
    \Line(0,5)(4,10)
    \Line(0,15)(4,10)
    \Line(20,5)(16,10)
    \Line(20,15)(16,10)
\end{axopicture}
\hspace{1pt}
}}

\begin{document}
\maketitle

\section{Introduction}
Feynman integrals \brk{FIs} are a corner stone of the perturbative
Quantum Field Theory \brk{pQFT}, at least in its contemporary formulation.
Within that branch of research one fruitful discovery were the
integration-by-parts identities \brk{IBPs}~\cite{Tkachov:1981wb,
Chetyrkin:1981qh}, i.e. linear identities among FIs.
For a given pQFT problem \brk{such as, for example, scattering amplitudes}
IBPs allow to reduce an infinite set of contributing FIs to a linear
combination of finite number of basic objects known as the master integrals
\brk{MIs}.

Recently a novel framework~\cite{
    Mastrolia:2018uzb,
    Frellesvig:2019kgj,
    Mizera:2019gea,
    Frellesvig:2019uqt,
    Frellesvig:2020qot%
} based on the twisted cohomology theory \cite{
    matsumoto:1994,
    cho:1995,
    matsumoto:1998,
    ohara:1998,
    OST:2003,
    % aomoto:2011,
    goto:2013,
    % yoshida:2013,
    goto:2015,
    goto:2015b,
    goto:2015c,
    Mizera:2017rqa,
    matsubaraheo:2019%
} was proposed to describe relations among FIs.
It was shown that \brk{for fixed topology of the corresponding Feynman graphs}
FIs form a finite dimensional vector space endowed with a scalar product called
the intersection number. Among other applications, this structure then helped
to derive novel algorithms for the direct projection of FIs onto the basis of MIs.

In \secref{sec:cohom} review the basics of twisted cohomology theory and
computation of intersection numbers.
Then in \secref{sec:second} we present another algorithm\footnote{
    This section is based on the joint work with
    Federico Gasparotto,
    Manoj K. Mandal,
    Pierpaolo Mastrolia,
    Saiei J. Matsubara-Heo,
    Henrik J. Munch,
    Nobuki Takayama.
}~\cite{Chestnov:2022alh}
for reduction of FIs exploiting the connection with the
Gel'fand-Kapranov-Zelevinsky \brk{GKZ} hypergeometric systems and the secondary
equation~\cite{matsubaraheo:2019}.

\section{Twisted cohomology}
\label{sec:cohom}
Here we review some aspects of the twisted cohomology and intersection theory
see also \cite{Mastrolia:2022tww, Mizera:2019ose, Frellesvig:2021vem,
Mandal:2022vok} and \cite{Weinzierl:2022eaz}.
Our central subject of study is going to be generalized hypergeometric
integrals of the form:
\begin{align}
    \FI = \int_\contour \twist\brk{x} \, \ccl\brk{x}
    \ ,
    % \equiv \int_\contour \twist\brk{x} \, \hat{\ccl}\brk{x} \cdot
    % \dd x_1 \wedge \ldots \wedge \dd x_n
    \label{eq:FI-def}
\end{align}
where $\twist\brk{x} = \prod_i \Baikov_i^{\gamma_i}$ is a multivalued function,
$\contour$ is an $n$-dimensional integration contour such that $\prod_i
\Baikov_i\brk{\partial \contour} = 0$, and $\ccl \equiv \hat{\ccl} \> \dd x_1
\wedge \ldots \wedge \dd x_n$ is a holomorphic $n$-form \brk{meaning that the
coefficient $\hat{\ccl}\brk{x}$ is a rational function}.

Integrals such as~\eqref{eq:FI-def} often appear as parametric representation of FIs. For
example, the Baikov representation of a FI with $\Loops$ loops and $E$
external legs in $\dims$ dimensions, the multivalued function $u\brk{x}$
contains a single factor $u\brk{x} = \Baikov\brk{x}^\gamma$, where $\Baikov$ is
the Baikov polynomial~\cite{Baikov:1996iu}, and the exponent
$\gamma = \brk{\dims - \Loops - E - 1} / 2$~. Hence in the following we will
refer to the integrals~\eqref{eq:FI-def} as generalized Feynman Integrals
\brk{GFI}.

Linear equivalence relation between FIs:
\begin{align}
    \FI =
    \int_\contour \twist \> \ccl \equiv \int_\contour \twist \> \brk{\ccl
    + \nabla_\omega \xi}\ ,
    \label{eq:FI-equiv}
\end{align}
where we introduced the covariant derivative: $\nabla_\omega \defas \dd +
\omega \wedge$ and the $1$-form
\begin{align}
    \omega \defas \dd \log{u}\ ,
    \label{eq:omega-def}
\end{align}
which will be very useful in the following.
%
% The equivalence relation~\eqref{eq:FI-equiv} follows from the properties of the
% $\twist\brk{x}$ function and the contour $\contour$, combined  with the Stokes theorem:
The equivalence relation~\eqref{eq:FI-equiv} follows from the Stokes theorem:
$
    0
    = \int_{\partial \contour} \twist \> \xi
    % = \int_\contour \dd\brk{\twist \> \xi}
    % = \int_\contour \twist \> \nabla_\omega \xi\ ,
    = \int_\contour \twist \> \nabla_\omega \xi
    % \underbrace{
    % }_{\nabla_\omega \xi}
    \ ,
$
% It shows that a generalized FI of a total covariant derivative
% $\nabla_\omega \xi$ vanishes.
    where $\nabla_\omega \xi \defas \dd \xi + \omega \wedge \xi$ is the
    covariant derivative.

Fixing the contour of integration $\contour$ allows us to interpret
relation~\eqref{eq:FI-equiv} as an equivalence of integrands.
Namely, we collect $n$-forms $\ccl$ into equivalence classes
$\bra{\ccl}: \ccl \sim \ccl + \nabla_\omega \xi$ generated by adding
covariant derivatives of $\brk{n - 1}$-forms.
Their totality forms the twisted cohomology group:
\begin{align}
    \bra{\ccl} \in \cohom^n_\omega
    \defas
    \bigbrc{
        \text{$n$-forms $\ccl$} \> | \>
        \nabla_\omega \ccl = 0
    }
    \Big/
    \bigbrc{
        \nabla_\omega \xi
    }\ ,
    \label{eq:cohom-def}
\end{align}
which can be thought of as the space of linearly independent FIs \brk{of a
given topology}.

Analogously we can introduce the dual integrals $\FI^\vee$, whose definition
mimics~\eqref{eq:FI-def} up to $u \mapsto u^{-1}$ and $\nabla_\omega
\mapsto \nabla_{-\omega}$\ . Elements of the dual twisted
cohomology group will be denoted by kets $\ket{\psi}$.

% Equivalence class of dual forms: $\ket{\ccr}: \ccr \sim \ccr +
% \nabla_{-\omega} \xi$ $\leadsto$ right forms.

% Dual twisted cohomology groups:
% \begin{align}
%     \ket{\ccr} \in \cohom^n_{-\omega}
%     \defas
%     \brc{
%         \text{$n$ forms} \> | \>
%         \nabla_{-\omega} \ccr_n = 0
%     }
%     /
%     \brc{
%         \nabla_{-\omega} \ccr_{n - 1}
%     }\ .
% \end{align}

\subsection{Counting the number of Master integrals}
The framework of twisted cohomology unites several seemingly independent
methods for computation of the number of MIs $\rank$:
\begin{enumerate}
    \item Number of unreduced integrals produced by the Laporta algorithm \cite{Laporta:2000dsw}.
    \item Number of critical points, i.e. solutions of $\dlog \twist\brk{x} = 0$\ \cite{Baikov:2005nv, Lee:2013hzt}.
    \item Number of independent integration contours $\contour_\lambda$ \cite{Bosma:2017ens, Primo:2017ipr}.
    \item Number of independent $n$-forms, i.e. $\Dim\bigbrk{\cohom^n_{\pm \omega}}$
        \cite{Mastrolia:2018uzb, Frellesvig:2020qot}.
    \item Holonomic rank of GKZ system \brk{volumes of certain polytopes}
        \cite{Chestnov:2022alh, Henrik:2022}.
\end{enumerate}

\subsection{Scalar product between Feynman integrals}
The twisted cohomology theory allows us to view the set of FIs
\brk{of a given topology} as a finite dimensional vector space.
%
% \begin{align}
%     &\FI = \spab{\ccl | \contour}
%     = \int_{\contour} \twist \> \ccl
%     \ ,\quad
%     \FI^{\vee} = \spba{\contour | \ccr}
%     = \int_{\contour} \twist^{-1} \> \ccr
%     % \ ,
%     % \\
%     % &\spab{c_1 \ccl_1 + c_2 \ccl_2 | \contour}
%     % = c_1 \spab{\ccl_1 | \contour} + c_2 \spab{\ccl_2 | \contour}
%     \ .
% \end{align}
%
A set of MIs $\bra{\lbasis_\lambda}$ for $\lambda \in \brc{1, \ldots, \rank}$
% $
%     \bigbrc{
%         \bra{\lbasis_\lambda}
%         % \text{for $\lambda \in \brc{1, \ldots, \rank}$}
%     }_{
%         \lambda \in \brc{1, \ldots, \rank}
%     }
% $
then forms a basis in that space.
%
% \begin{align}
%     \bigbrc{
%         \bra{\lbasis_\lambda} \>\big|\>
%         \text{for $\lambda \in \brc{1, \ldots, \rank}$}
%     }
%     \ .
% \end{align}

The dual FIs really form a dual vector space to FIs due to the
existence of a scalar product:
\begin{align}
    \vev{\ccl | \ccr}
    = \frac{1}{\brk{2 \pi \imi}^n}
    \int \iota\brk{\ccl} \wedge \ccr
    \ ,
    \label{eq:interx-def}
\end{align}
called the intersection number.
This scalar product allows to directly
decompose a given integral $\FI$
in a basis of MIs $\cJ_\lambda \defas \int_\contour \twist \> \lbasis_\lambda$,
namely $\FI = \sum_{\lambda = 1}^\rank c_\lambda \, \cJ_\lambda$\ .
Linear algebra leads us to the master decomposition
formula~\cite{Mastrolia:2018uzb, Frellesvig:2020qot}:
\begin{align}
    &\bra{\ccl} = \sum_{\lambda = 1}^\rank c_\lambda \,
    \bra{\lbasis_\lambda}\ ,
    \quad
%\end{align}
%%
%where the decomposition coefficients are given by:
%%
%\begin{align}
    c_\lambda = \sum_{\mu = 1}^\rank \vev{\ccl | \rbasis_\mu}
    \bigbrk{\Cmat^{-1}}_{\mu \lambda}
    \ ,
    \\
    &\Cmat_{\lambda \mu} \defas \vev{\lbasis_\lambda |
    \rbasis_\mu}
    \ ,
    \label{eq:cmat-def}
\end{align}
for any choice of the dual basis $\ket{\rbasis_\mu}$.
Therefore the intersection numbers~\eqref{eq:interx-def} completely determine the
decomposition coefficients. Let's see now how they can be computed.

\subsection{Univariate intersection numbers}
\label{ssec:uni}
In the $n = 1$ case, intersection numbers~\eqref{eq:interx-def}
turn into a sum of residues~\cite{cho:1995, matsumoto:1998, Frellesvig:2021vem}:
\begin{align}
    \vev{\ccl | \ccr}
    % &= \frac{1}{2 \pi \imi} \int_X \iota\brk{\ccl} \wedge \ccr
    % \\
    \equiv \frac{1}{2 \pi \imi} \int_X \lrbrk{
        \ccl - \sum_{p \in \Poles_\omega}
        \nabla_\omega \bigbrk{\Heaviside_p\brk{x, \bar{x}} \potential_p}
    }\wedge \ccr
    = \sum_{p \in \Poles_\omega} \res{x = p}\lrsbrk{\potential_p \, \ccr}\ ,
\end{align}
where
\begin{itemize}
    \item Integration goes over $X = \cp^1$.
    % \item regularization $\iota\brk{\ccl}$\ .
    \item $\Poles_\omega \defas \bigbrc{p \>\big|\> \text{poles of $\omega$}}$, including
        the $\infty$\ .
    \item Terms with Heaviside $\theta$-functions regulate the integral with
        the help of a local potential $\potential_p$, which satisfies
        $\nabla_\omega \potential_p \equiv \brk{\dd + \omega \wedge} \potential_p = \ccl$
        around the pole $p$\ .
        This differential equation can be solved via an Ansatz: $\potential_p =
\potential_{p, \, \Min} \brk{x - p}^{\Min} +
\potential_{p, \, \Min + 1} \brk{x - p}^{\Min + 1} + \ldots +
\potential_{p, \, \Max} \brk{x - p}^{\Max}
$\ .
\end{itemize}

\subsection{Multivariate intersection numbers}
One strategy for dealing with the intersection numbers of multivariate FIs
is to apply the univariate procedure recursively one variable at a
time~\cite{ohara:1998, Mizera:2017rqa, Mastrolia:2018uzb, Frellesvig:2019uqt, Frellesvig:2020qot}.
Consider a 2 variable problem: given two 2-forms $\ccl\brk{x_1, x_2}$ and $\ccr\brk{x_1, x_2}$
we would like to compute $\vev{\ccl | \ccr}$ by first integrating out $x_1$ and then $x_2$\ .
To do that we pick a basis $\bra{\lbasis_\lambda}$ and its dual
$\ket{\rbasis_\mu}$ for the internal $x_1$-integration and project $\ccl$,
$\ccr$ onto them \brk{omitting the summation signs}:
\begin{alignat}{3}
    \bra{\ccl} &= \bra{\lbasis_\lambda} \wedge \bra{\ccl_{\lambda}}
    \ ,
    \quad
    &&
    \bra{\ccl_{\lambda}} = \vev{\ccl | \rbasis_\mu}
    \bigbrk{\Cmat^{-1}}_{\mu \lambda}
    \ ,
    &&
    \\
    \ket{\ccr} &= \ket{\rbasis_\mu} \wedge \ket{\ccr_{\mu}}
    \ ,
    \quad
    &&
    \ket{\ccr_{\mu}} = \bigbrk{\Cmat^{-1}}_{\mu \lambda}
    \vev{\lbasis_\lambda| \ccr}
    \ ,
    &&
    \label{eq:psi-proj}
\end{alignat}
The internal $x_1$-integration can be seen as the insertion of the identity operator
$
    \IdOp\subc
    = \ket{\rbasis_\mu} \bigbrk{\Cmat^{-1}}_{\mu \lambda}
    \bra{\lbasis_\lambda}\ ,
    % \label{eq:id-def}
$
which consequently allows us to write the remaining integral in $x_2$ as a sum over residues:
\begin{align}
    \vev{\ccl | \ccr} =
    \langle \ccl \underbrace{
        | \rbasis_\mu \rangle
        \bigbrk{\Cmat^{-1}}_{\mu \lambda}
    \langle \lbasis_\lambda |
    }_{\IdOp\subc}
    \ccr \rangle
    =
    \sum_{p \in \Poles_\Pfaff} \Res_{x_2 = p}\lrsbrk{
        \potential_{p, \lambda} \, C_{\lambda \mu} \, \ccr_{\mu}
    }
    \ .
    \label{eq:interx-res}
\end{align}
Similar to~\secref{ssec:uni}, this formula requires the knowledge of a local vector potential
$\potential_{p, \lambda}$ near each pole $x_2 = p$.
The potential is fixed by the following system of first order differential
equations \brk{omitting the $p$ subscript}:
\begin{align}
    \partial_{x_2} \potential_{\lambda} + \potential_{\mu} \> \Pfaff_{\mu
    \lambda} = \ccl_\lambda
    \quad \text{near $x_2 = p$}\ .
\end{align}
The differential equation matrix $\Pfaff$ and it's dual version $\Pfaff^\vee$
are made out of $x_1$-intersection numbers:
\begin{align}
    \Pfaff_{\lambda \nu} \defas
    \vev{
        \brk{\partial_{x_2} + \omega_2} \lbasis_\lambda | \rbasis_\mu
    }
    \bigbrk{\Cmat^{-1}}_{\mu \nu}
    \ ,\quad
    \Pfaff^\vee_{\mu \xi} \defas
    \bigbrk{\Cmat^{-1}}_{\mu \lambda}
    \vev{
        \lbasis_\lambda |
        \brk{\partial_{x_2} - \omega_2} \rbasis_\xi
    }
    \ ,
    \label{eq:pfaff-def}
\end{align}
so they can be computed using the univariate method of~\secref{ssec:uni}. The
set $\Poles_\Pfaff$ in eq.~\eqref{eq:interx-res} is defined as $\Poles_\Pfaff
\defas \bigbrc{p \> \big| \> \text{poles of $\Pfaff$}}$\ .

In practice, to compute the residue at, say, $x_2 = 0$ we solve for $\rho$ the
following system:
\begin{align}
    \begin{cases}
    % \begin{aligned}
        \lrsbrk{x_2 \, \partial_{x_2} + \Pfaff\brk{x_2}} \vv{\potential} = \vv{\ccl}
        \\[5pt]
        \rho = \res{x_2 = 0} \lrsbrk{
            \vv{\potential} \, \cdot \, \vv{\ccr}
        }
    % \end{aligned}
    \end{cases}
    \ ,
    \label{eq:res-sys}
\end{align}
%
% where we rescaled the matrix~\eqref{eq:pfaff-def} as $\Pfaff\brk{x_2} \mapsto 1
% / x_2 \> \Pfaff\brk{x_2}$,
where we rescaled $\vv{\ccl}\brk{x_2} \mapsto 1 / x_2 \> \vv{\ccl}\brk{x_2}$ and
$\Pfaff\brk{x_2} \mapsto 1 / x_2 \> \Pfaff\brk{x_2}$,
and canceled the $\Cmat$ matrix in the residue~\eqref{eq:interx-res} against the
$\Cmat^{-1}$ coming from eq.~\eqref{eq:psi-proj}.
The series expansion of the system~\eqref{eq:res-sys} is build from:
\begin{align}
% $
    \Pfaff\brk{x_2} = \sum_{i \ge 0} x_2^i \> \Pfaff_i
    \ , \quad
    \vv{\ccl} = \sum_{i \ge k} x_2^i \> \vv{\ccl}_i
    \ , \quad
    \vv{\ccr} = \sum_{i \ge m} x_2^i \> \vv{\ccr}_i
    \ ,
    % $
\end{align}
% \begin{align}
%     \Pfaff\brk{x_2} = \sum_i x_2^i \> \Pfaff_i = \Pfaff_0 + x_2 \> \Pfaff_1 +
%     \ldots
%     \ , \quad
%     \vv{\ccl} = x_2^{-k} \vv{\ccl}_{-k} + \ldots
%     \ , \quad
%     \vv{\ccr} = x_2^{-m} \vv{\ccl}_{-m} + \ldots
%     \ ,
% \end{align}
%
for integer $k, m \in \Integers$.
Inserting an Ansatz
$\vv{\potential} = \sum_i x_2^i \> \vv{\potential}_i$\ ,
and matching the powers of $x_2$ order by order, we obtain the linear system
\brk{here dots $\mzero$ denote zeros}:
\begin{align}
    \lrsbrk{
        \begin{array}{c|ccccc|c}
            -1
            & \vv{\ccr}_{1} & \vv{\ccr}_{0} & \vv{\ccr}_{-1} & \vv{\ccr}_{-2} & \vv{\ccr}_{-3}
            % & \cellcolor{gr1}\tikzmark{end}\mzero
            & \cellcolor{gr1}\mzero
            \\
            \mzero
            & \Pfaff_0 - 2 & \mzero & \mzero & \mzero & \mzero
            & \vv{\ccl}_{-2}
            \\
            \mzero
            & \gr{\Pfaff_1} & \Pfaff_0 - 1 & \mzero & \mzero & \mzero
            & \vv{\ccl}_{-1}
            \\
            \mzero
            & \gr{\Pfaff_2} & \gr{\Pfaff_1} & \Pfaff_0 & \mzero & \mzero
            & \vv{\ccl}_{0}
            \\
            \mzero
            & \gr{\Pfaff_3} & \gr{\Pfaff_2} & \gr{\Pfaff_1} & \Pfaff_0 + 1 & \mzero
            & \vv{\ccl}_{1}
            \\
            \mzero
            & \gr{\Pfaff_4} & \gr{\Pfaff_3} & \gr{\Pfaff_2} & \gr{\Pfaff_1} & \Pfaff_0 + 2
            & \vv{\ccl}_{2}
        \end{array}
    }
    \cdot
    \lrsbrk{
        \def\arraystretch{1.2}
        \begin{array}{c}
            \rho
            \\
            \vv{\potential}_{-2}
            \\
            \vdots
            \\
            \vv{\potential}_2
            \\
            -1
        \end{array}
    } = 0
    \ .
\end{align}
This equation has to be solved only for $\rho$\ .
Row reduction of this matrix can be carried out only until the first row is
filled with zeros except for the element in the last column \brk{highlighted
with grey}, which will contain the needed residue.
Other poles of eq.~\eqref{eq:interx-res} are treated in the same manner and
the sum of their residues produces the intersection number $\vev{\ccl | \ccr}$\ .
% We repeat the same procedure for all the other poles $p \in \Poles_\Pfaff$ of
% eq.~\eqref{eq:interx-res}, sum everything together and produce the value of the
% intersection number $\vev{\ccl | \ccr}$.
% Resonances.
% Higher order poles.
% If $\Pfaff$ is ``simple pole'' $\leadsto$ Global residue theorem. \cite{Weinzierl:2020xyy}

\section{Decomposition via the secondary equation}
\label{sec:second}
As was observed in \cite{Chestnov:2022alh} \brk{see also \cite{Henrik:2022}},
the twisted cohomology framework provides another method for computation of
the decomposition coefficients~\eqref{eq:cmat-def}.
The first key idea is the so-called secondary equation
\cite{matsubaraheo:2019, Frellesvig:2020qot, Weinzierl:2020xyy}, which is a
matrix differential equation satisfied by the intersection matrix
$\Cmat$:
% A way to decompose $\ccl = \sum_{\lambda = 1}^\rank c_\lambda \>
% \lbasis_\lambda$
% % $c_\lambda = \sum_{\mu = 1}^\rank \vev{\ccl | \rbasis_\mu} \bigbrk{\Cmat^{-1}}_{\mu \lambda}$
% via $\Cmat_{\lambda \mu} = \vev{\lbasis_\lambda | \rbasis_\mu}$\ .
% \end{align}
\begin{align}
    \begin{cases}
        \partial_{z_i} \, \bra{\lbasis_\lambda} = \bigbrk{\Pfaff_i}_{\lambda \nu}
        \, \bra{\lbasis_\nu}
        \\
        \partial_{z_i} \, \ket{\rbasis_\mu} = \ket{\rbasis_\xi} \, \bigbrk{\Pfaff^\vee_i}_{\xi \mu}
    \end{cases}
    \Longrightarrow
    \partial_{z_i} \, \Cmat =
    \Pfaff_i \cdot \Cmat + \Cmat \cdot \lrbrk{\Pfaff_i^\vee}\transpose
    \ ,
    \label{eq:second}
\end{align}
where $z_i$ are some external kinematical variables.
The other key step is computation of the differential equation
matrices $\Pfaff$ and $\Pfaff\aux$ made available thanks to the connection
of the  twisted cohomology theory, the GKZ formalism, and $D$-module theory.
We assume that this step is completed and refer the
interested reader to~\cite{Chestnov:2022alh, Henrik:2022} for the full story.
Once the secondary equation~\eqref{eq:second} is written down, we employ the known
algorithms for finding rational solutions of such systems, e.g.
the \soft{Maple} package \soft{IntegrableConnections}~\cite{Barkatou:2012}.

Finally, to determine the decomposition coefficients~\eqref{eq:cmat-def} we
repeat the above procedure for an auxiliary basis $\lbasis\aux \defas
\brc{\lbasis_1, \ldots, \lbasis_{\rank - 1}, \ccl}$, i.e. we compute an
auxiliary $\Pfaff\aux$ and then $\Cmat\aux$\ . The FI decomposition is then
encoded in the following matrix product:
\begin{align}
    \lrsbrk{
        % \begin{small}
            \arraycolsep = -1pt \def \arraystretch{0.9}
            \begin{array}{c}
                \lbasis_1\\
                \vdots\\
                \lbasis_{\rank-1}\\
                \ccl
            \end{array}
        % \end{small}
    }
    =
    \Cmat\aux \cdot \Cmat^{-1}
    \lrsbrk{
        % \begin{small}
            \arraycolsep = -1pt \def \arraystretch{0.9}
            \begin{array}{c}
                \lbasis_1\\
                \vdots\\
                \lbasis_{\rank-1}\\
                \lbasis_\rank
            \end{array}
        % \end{small}
    }
    \quad
    \Longrightarrow
    \quad
    \Cmat\aux \cdot \Cmat^{-1}
    = \lrsbrk{
        % \begin{small}
            \arraycolsep = .7pt \def \arraystretch{0.9}
            \begin{array}{ccc|c}
                & & & 0
                \\
                & {\Id_{\rank-1}} & & \vdots
                \\
                & & & 0
                \\
                \hline
                \rowcolor{gr1}
                c_1 & \cdots & c_{\rank-1} & c_\rank
            \end{array}
        % \end{small}
    }
    \ ,
    \label{eq:CauxCinv}
\end{align}
where $\Id_{\rank - 1}$ denotes an identity matrix of size $\brk{\rank - 1}$,
and the decomposition coefficients $c_\lambda$ are collected in the last row
highlighted with grey.

\subsection{A simple example}
Let us briefly showcase how the secondary equation can produce the reduction
coefficients of a box diagram with a single dot $\ccl = \boxd$ in terms of the
basis $
    \brk{\lbasis_1, \lbasis_2, \lbasis_3} = \lrbrk{
        \tbub,
        \sbub,
        \boxx\
    }
$\ .
This topology has
$\twist = \brk{x_1 + x_2 + x_3 + x_4 + x_1 x_3 + t / s \> x_2 x_4}^\gamma$\ .
Using the algorithm of
\cite{Chestnov:2022alh} and the \soft{Asir} computer algebra system
\cite{url-asir} we generate the differential equation matrices:
\begin{align}
    P = \lrsbrk{
    \arraycolsep = 2pt \def \arraystretch{1.5}
    \begin{array}{ccc}
     - \frac{ \epsilon \left(\delta ^2 (12 z+11)+7 \delta  (z+1)+z+1 \right)}{(3 \delta
       +1) z (z+1)} & -\frac{\delta ^2 \epsilon }{(3 \delta+1)(z+1)} & \frac{\delta ^2
       \epsilon  (\delta  (z+2)+1)}{2 (3 \delta +1) z
       (z+1) (\delta  \epsilon +1)} \\
     \frac{\delta ^2 \epsilon }{(3 \delta +1)
       z \left(z+1\right)} & -\frac{\delta ^2
       \epsilon }{(3\delta+1) (z+1)} &
       -\frac{\delta ^2 \epsilon  (\delta +2 \delta
       z+z)}{2 (3 \delta +1) z (z+1) (\delta
       \epsilon +1)} \\
     -\frac{2 (2 \delta +1) \epsilon  (\delta
       \epsilon +1)}{(3 \delta +1) z (z+1)} & \frac{2
       (2 \delta +1) \epsilon  (\delta  \epsilon
       +1)}{(3 \delta +1) (z+1)} & -\frac{\epsilon
       \left(\delta ^2 (5 z+7)+\delta  (2
       z+5)+1\right)}{(3 \delta +1) z (z+1)}
    \end{array}
    }\ ,
\end{align}
where $z = t / s$ is the ratio of the Mandelstam invariants, $\delta$ is an
additional regularization parameter which should be set $\delta \to 0$ at the
end of the computation, and $\Pfaff^\vee = \Pfaff \big|_{\epsilon \to
-\epsilon}$
\brk{see~\cite{Chestnov:2022alh,Henrik:2022} for further details}.
The rational solution to the secondary equation~\eqref{eq:second}
% \brk{found using the \soft{IntegrableConnections} package}
looks like this:
\begin{align}
    \Cmat =
    \lrsbrk{
    \arraycolsep = 1pt \def \arraystretch{1}
    \begin{array}{ccc}
     -\frac{(2 \delta +1) (4 \delta +1)}{\delta } &
       \delta  & -2 (\delta  \epsilon -1) \\
     \delta  & -\frac{(2 \delta +1) (4 \delta
       +1)}{\delta } & -2 (\delta  \epsilon -1) \\
     2 (\delta  \epsilon +1) & 2 (\delta  \epsilon
       +1) & -\frac{4 \left(10 \delta ^2+6 \delta
       +1\right) (\delta  \epsilon -1) (\delta
       \epsilon +1)}{\delta ^3}
    \end{array}
    }
    \ .
\end{align}
We repeat the same procedure for the auxiliary $\Cmat\aux$ and mount everything
into eq.~\eqref{eq:CauxCinv} to produce:
\begin{align}
    \boxd
    =
    -\frac{2 \eps \brk{2 \eps + 1}}{z \brk{\eps + 1}} \cdot
    \tbub
    + 0 \cdot
    \sbub
    + \brk{2 \eps + 1} \cdot
    \boxx
    \ .
\end{align}
Therefore the secondary equation method allows us to decompose FIs in terms of MIs via
solving a first order matrix differential equation~\cite{Chestnov:2022alh}!

\section{Conclusion}
We reviewed the connection between FIs and the twisted cohomology theory,
focusing on the algorithms for computation of the uni- and multivariate
intersection numbers, that is the scalar products between FIs.

Furthermore, following~\cite{Chestnov:2022alh}, we showed how the twisted
cohomology together with the theory of GKZ hypergeometric system provide a
way to compute the IBP reduction coefficients via essentially finding rational
solutions to a system of PDEs~\eqref{eq:second} called the secondary equation.
In the future it would be interesting to further develop this connection and
apply it to other problems and processes within pQFT.

Figures were made with \soft{AxoDraw2}~\cite{Collins:2016aya}.

\bibliographystyle{JHEP}
\bibliography{references}

\end{document}